# From Wye-Delta to Cross-Square Recursion Configurations in Graphene-Based Quantum Hall Arrays

N. T. M. Tran,[1,2] M. Musso,[3] D. S. Scaletta,[4] W.-C. Lin,[5] V. Ortiz Jimenez,[1] D. G. Jarrett,[1] M. Ortolano,[3] C. A. Richter,[1] C.-T. Liang,[6] D. B. Newell,[1] and A. F. Rigosi[1,a]

[1]*Physical Measurement Laboratory, National Institute of Standards and Technology (NIST), Gaithersburg, Maryland 20899, USA*

[2]*Joint Quantum Institute, University of Maryland, College Park, Maryland 20742, USA*

[3]*Department of Electronics and Telecommunications, Politecnico di Torino, Torino 10129, Italy*

[4]*Department of Physics, Mount San Jacinto College, Menifee, California 92584, USA*

[5]*Department of Engineering and System Science, National Tsing Hua University, Hsinchu 300044, Taiwan*

[6]*Department of Physics, National Taiwan University, Taipei 106053, Taiwan*

In electrical metrology, the quantum Hall effect is accessed at the Landau level filling factor $\nu = 2$ plateau to define and disseminate the unit of electrical resistance (ohm). The robustness of the plateau is only exhibited at this Landau level filling factor and thus places a constraint on the quantized resistances that are accessible when constructing quantized Hall array resistance standards (QHARS) using epitaxial graphene on SiC. To overcome devices constrained by using Hall elements in series or in parallel, this work approaches the fabrication of a cross-square network configuration, which is similar to but departs slightly from conventional wye-delta designs and achieves significantly higher effective quantized resistance outputs. Furthermore, the use of pseudofractal-like recursion amplifies the ability to reach high resistances. QHARS devices designed as the ones here are shown to achieve an effective resistance of 55.81 MΩ in one configuration and 27.61 GΩ in another, with a hypothetically projected 317.95 TΩ that could be accessed with more specialized equipment. Teraohmmeter measurements reveal the limits of conventional wet cryogenic systems due to resistance leakage. Ultimately, this work builds on the capability of realizing exceptionally high-value quantum resistance standards.

______________________

[a] Author to whom correspondence should be addressed.  afr1@nist.gov

In recent years, epitaxial graphene (EG)-based resistance standards on SiC have been fabricated with the intent of making more quantized resistances available for metrological applications [1-7]. In the context of the history of the International System of Units (SI), the ohm has been most pragmatically disseminated by measuring a single plateau of the quantum Hall effect ($v = 2$; $\frac{1}{2}\frac{h}{e^2} = \frac{1}{2}R_K \approx 12906.4037\ \Omega$, where $h$ is the Planck constant and $e$ is the elementary charge), and having only one accessible value greatly burdens the measurement infrastructure necessary to disseminate the ohm. Therefore, recent efforts to access different quantized values include approaches to build quantized Hall array resistance standards (QHARS), which entail the assembly of multiple Hall bars in parallel, series, or arranged as *p-n* junctions to access resistances valued at $qR_K$, where $q$ is a positive rational number [8-19]. Building QHARS devices then becomes limited mostly by the total area over which high-quality EG may be grown since each element has a finite minimum size, with that constraint yielding a total number of Hall elements that can be fabricated onto a chip.

The highest resistances achievable by placing QHARS elements in series (assuming a rough estimate of 1000 elements) are on the order of 10 M$\Omega$ [20-21], which is many orders of magnitude under the full scale of metrology needs, with some going as high as P$\Omega$ levels [22]. To circumvent this problem of scaling to higher resistances, QHARS devices may be designed as wye-delta (Y-$\Delta$) networks [23-24]. To go beyond this order of 10 M$\Omega$ more easily, one can adopt networks with more grounded branches, which in the general case is known as the star-mesh transformation and has been explored primarily on a theoretical basis, and has been demonstrated experimentally, achieving effective quantized resistances as high as 1 G$\Omega$ [25-29]. In this work, a departure from the classic wye-delta network (to a cross-square configuration) shows that high effective quantized resistances are more easily accessible when using more than one grounded branch. Furthermore, measurements of the device when wirebonded in ways that correspond to additional cross-square recursions reveal the limits of conventional cryogenic systems when using standard lock-in techniques and self-calibrating teraohmmeters (T$\Omega$ meter) [30].

For device fabrication, 4H-SiC chips were cleaned with a commercially stabilized mixture of sulfuric acid and hydrogen peroxide and the afterward with hydrofluoric acid. Prior to growth, they were coated with a carbon-based photoresist (AZ 5214E) and pressed against glassy carbon to ensure close spacing and limit Si escape, improving graphene uniformity [31]. A growth furnace was heated to 1850 °C for 3 min to 4 min to form EG on a carbon buffer layer. The EG films were then characterized by optical and confocal laser scanning microscopy to select those lateral areas with over 99 % monolayer coverage as opposed to bilayer or buffer layer EG [31]. For device fabrication, the EG layer was protected by a Pd/Au layer (not greater than 100 nm), followed by etching and photolithography to define the Hall bar and contact patterns.

Superconducting NbTiN was deposited over the protective Pd/Au layer in the contacting region to form interconnects, and it is separated from the EG by more than 80 nm to prevent quantum effects like Andreev reflection [10, 12]. Gateless control of carrier density was achieved by functionalizing the EG with Cr(CO)$_3$ in a nitrogen-filled furnace at 130 °C, ensuring uniform carrier density [32-33]. The usual value for the carrier density after ambient atmospheric exposure is close to the Dirac point of graphene, namely around $10^{10}$ cm$^{-2}$ [33], and this may be compared with other values of inherent doping in EG of $10^{13}$ cm$^{-2}$ [34].

When it comes to the layout of the QHARS device, it was important to consider diagnostics in the event of poor quantization within a subarray of the device. As shown in Fig. 1 (a), large contact pads were fabricated at the end of every row (spanning 10 Hall elements), allowing access to quantized values in multiples of 129 kΩ. This design was also implemented for purposes of access at least 10 different values on the order of kΩ, as described in Ref. [35]. As per the mathematical framework used to optimize device designs using pseudofractal recursions [28-29], it remained convenient to select a number of Hall elements that would be evenly symmetric with each iteration for multiple recursions. As with earlier forms of wye-delta arrays, it is optimal to constrain the grounded branch to be a single Hall element, and the key difference in this case is to increase the grounded branch count by one, resulting in a slightly modified wye-delta transformation (but not as general as the star-mesh case), as visualized in Fig. 1 (a):

$$R_{ij} = \frac{R_i R_j}{R_i} + \frac{R_i R_j}{R_j} + \frac{R_i R_j}{R_k} + \frac{R_i R_j}{R_l}$$

(1)

In Eq. 1, the condition is that $i \neq j$. To predict the output of an optimally designed QHARS device, let us define $q \equiv \frac{R}{R_H}$, where $q$ is the number of single Hall elements held at EG's $\nu = 2$ quantum Hall plateau to obtain the total resistance $R$. Note that this coefficient $q$, the *coefficient of effective resistance* (CER), is restricted to the set of positive integers ($q: q \in \mathbb{Z}^+$). And with $q_k = q_l$, Eq. 1 then becomes:

$$q_{ij} = q_i + q_j + \frac{q_i q_j}{q_k} + \frac{q_i q_j}{q_l} = q_i + q_j + 2\frac{q_i q_j}{q_k}$$

(2)

Recall from Ref. [28] that the parameter *M*, or recursion number, was simply the number of iterations of a pseudofractal, as shown in the bottom panels of Fig. 1 (b). Other parameters included: $q_{M:i}$ (single index, actual number of Hall elements) and $q_{M:ij}^{(approx)}$ (two indices, the *effective* number of elements). For the cross-square configuration, one obtains:

$$q_{M:i} = \frac{1}{2}\left(2q_{M:ij}^{(approx)} + 1\right)^{2^{-M}} - \frac{1}{2}$$

(3)

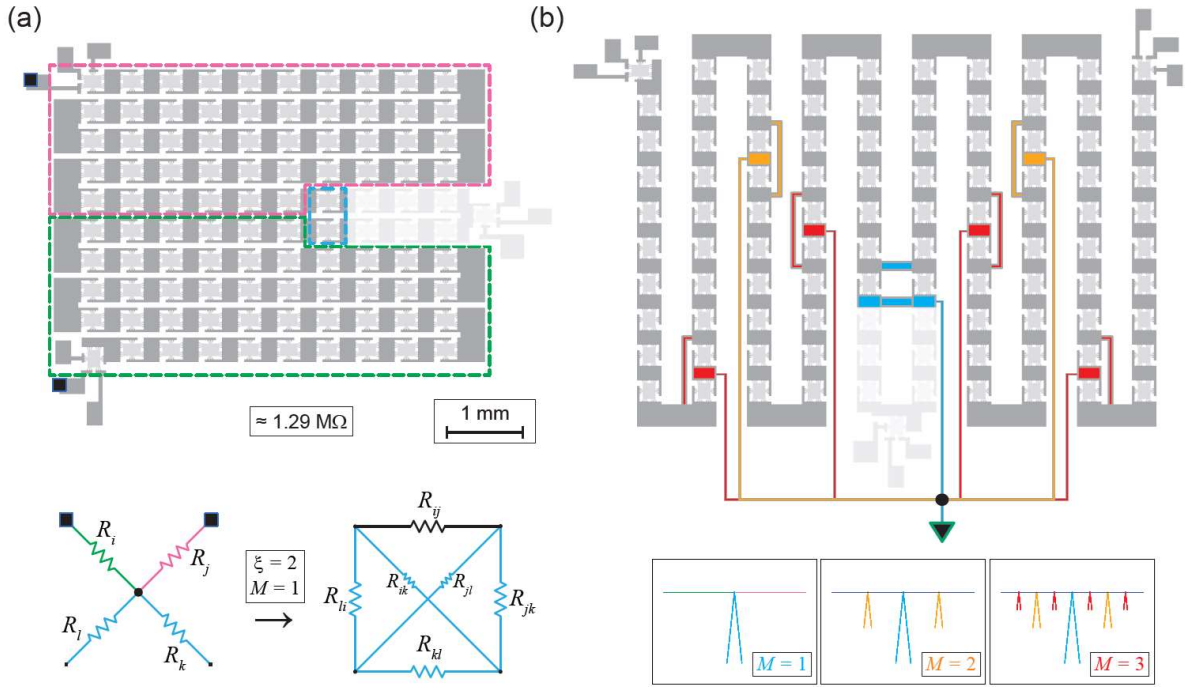

FIG. 1. (a) Illustration of the exact design of a 1.29 MΩ QHARS device is shown with two spots marked with squares indicating the start and end of the cross-square network. Two subarrays are outlined by dashed green and pink lines and match the resistances in the drawing below. Part of the array is outlined by a dashed blue line indicating the grounded branches. For device symmetry during recursion, a subset of the array is excluded from the circuit (faded area to the right of the dashed blue box). (b) A modified diagram of the QHARS device design is shown to illustrate how the wirebonding plan is coupled to the recursion factor *M*. The three diagrams at the bottom are drawings that represent the topology of the device. Each line represents a quantized resistance (whose value depends on the corresponding circuit). For example, in the *M* = 1 diagram, the two horizontal lines (in green and pink) represent the two subarray quantized resistances in (a) and each of the two light blue lines represents a grounded branch, each set to one quantum Hall element. As an additional example, the *M* = 2 diagram contains four horizontal segments that are each representing 22 quantum Hall elements in series (and the additional grounded branches in orange are made smaller and have a color change for two respective purposes: (1) to guide the eye in visualizing the second recursion and (2) to see what wiring is applied to achieve the second recursion).

Also, from Ref. [28], one can derive the cross-square configuration's total number of Hall elements in an optimal QHARS device:

.

$$D_T(M, q_{M:ij}) = 2^{M-1}\left(2q_{M:ij}^{(approx)} + 1\right)^{2^{-M}} - 2^{M-1} + 2(2^M - 1)$$

(4)

Using Eq. 4, one can essentially predict the expected values of the QHARS device output. Though the full device, when measured in series, yields about 1.29 M$\Omega$, strategic wirebonding and grounding may allow one to go well beyond the M$\Omega$ level. Figure 1 (b) shows a set of color-coded wires that utilizes cross-square transforms on the QHARS device. In the first case, $M = 1$, only two pairs of centrally located wirebonds are implemented (in light blue). The top pair of linked elements keeps the left and right halves of the device fully connected, whereas the bottom pair of linked elements serves as the two grounded single-element branches ($R_k$ and $R_l$) per the drawing in Fig. 1 (a). Knowing $D_T$ (94 total elements) and $M$, one may calculate that this cross-square configuration will yield about 55.81 M$\Omega$ when measured across the two main terminals (black squares in Fig. 1 (a)). If a second iteration of the cross-square configuration is implemented (as per the central lower inset of Fig. 1 (b)), the predicted value of the QHARS output becomes 27.61 G$\Omega$. This implementation keeps the light blue and introduces the orange illustrated wiring. And lastly, if the red-colored wiring is implemented (for the $M = 3$ recursion), then the predicted value becomes 317.95 T$\Omega$. These three values were confirmed with LTspice simulations [18, 36-37].

With the output for each of the cross-square recursions known, measurements could be taken and compared. QHARS device transport properties were assessed with a Janis Cryogenics $^4$He cryostat (see Acknowledgments). All devices were mounted onto a transistor outline (TO-8) package, and all corresponding magnetoresistance data were collected between magnetic field values of 0 T and ± 9 T. All measurements were performed at approximately 1.5 K with source-drain currents at 100 nA when using a lock-in amplifier or lower when using the self-calibrating T$\Omega$ meter. Prior to cooldown, devices were annealed in vacuum as described in Ref. [33] to obtain a desired electron density corresponding to a $\nu = 2$ plateau onset of approximately 3 T.

The next steps were to validate both the measuring equipment and the quantization of the QHARS device prior to measuring the cross-square configurations. Using lock-in amplifiers, the magnetoresistance of the QHARS device subarrays were measured as a function of magnetic field and plotted in Fig. 2 (a) and two blue squares indicate the source and drain. The two halves of the device (colored in green and pink) are measured separately and using the central 47$^{th}$ element as a drain (of which there are two in light blue – see inset of the same panel). While accounting for the input impedance of the circuit (which is mainly the 10 M$\Omega$ impedance from the lock-in amplifier), the two halves show identical quantization at about 614.5 k$\Omega$ (with an expected value of about 606.6 k$\Omega$). A second method to test parts of the full array utilizes different grounding branches in a way that allows for a two-terminal measurement of the first and last 11-element legs, which was expected to be about 141.9 kΩ and measured to be about 140.1 kΩ (see Fig. 2 (b) and its inset for the corresponding illustration). Both the QHARS quantization and the precision of the lock-in amplifier were suitable for testing the lower (MΩ level) resistances, but for those much higher than the impedance of the amplifier, a Guildline 6500A TΩ self-calibrating meter was used and needed validation (see Acknowledgments).

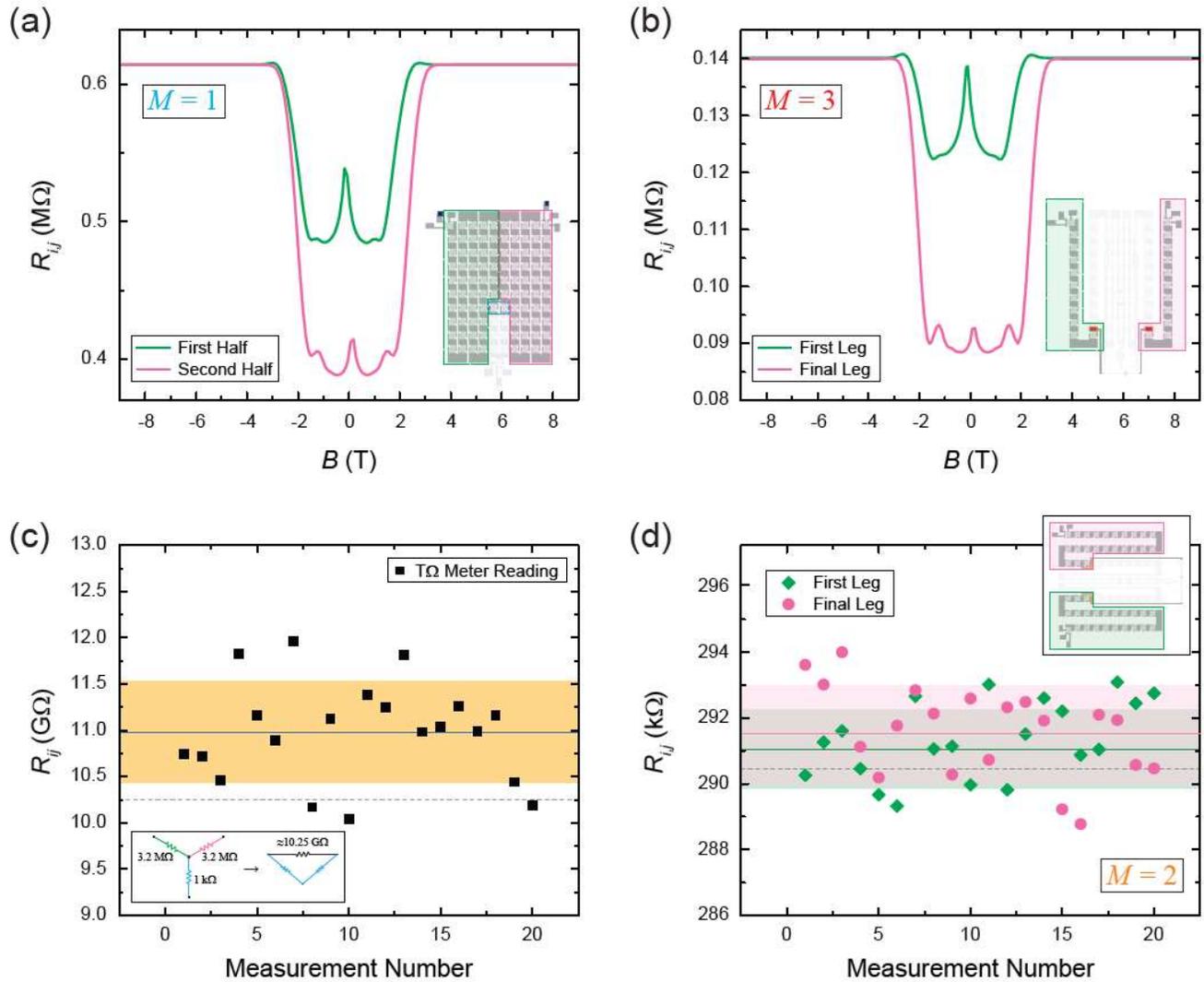

FIG. 2. (a) The magnetoresistance of QHARS device subarrays, where the two halves of the device (colored in green and pink) are measured with a lock-in amplifier and account for the input impedance of the circuit. These measurements are meant to validate the quantization of the subarrays prior to use in the cross-square configuration. (b) Two-terminal, lock-in amplifier magnetoresistance measurements were taken on the first and final legs of the QHARS device after being wirebonded for the $M = 3$ iteration of the cross-square configuration. (c) A test resistor Wye-delta network was used for confirming the accuracy and precision of a self-calibrating teraohmmeter (TΩ meter) used for later high-resistance measurements. (d) Verification of QHARS device subarray values based on an $M = 2$ iteration of the cross-square configuration per the inset illustration. Each measurement for (c) and (d) takes approximately 12 s, with the solid lines indicating the average of the measurement set and the shaded areas indicating the standard uncertainty.

Specifically, the TΩ meter precisely measures high resistance values by using a sophisticated constant-voltage source and a sensitive current-sensing electrometer. The instrument applies a user-selectable, highly stable direct current voltage across the material, while simultaneously measuring the small current that flows through it. This measurement is achieved using a low-noise, high-impedance input stage that minimizes measurement uncertainty. To accomplish the validation of this equipment relative to the order of magnitude of the measurements, an example wye-delta resistor network was used. Two of the three arms of the network were valued at 3.2 MΩ (based on commercially available resistors), with the last arm valued at 1 kΩ. The final calculated effective resistance was about 10.25 GΩ and was compared to the series of measurements taken with the TΩ meter in Fig. 2 (c), with each measurement of the TΩ meter taking approximately 12 s. The dotted gray line indicates this calculated value, and the navy line describes the average of the full measurement campaign, along with the gold shaded area being the standard uncertainty.

A TΩ meter test was also performed on the first and final legs of the QHARS device while wired for the $M = 2$ cross-square recursion, as shown in Fig. 2 (d). Since each of these legs contained 22.5 elements in series (when carefully considering the wiring of the final element), the predicted value was 290.4 kΩ. As seen in the panel, the first and final legs were measured to be about 291.1 kΩ and 291.4 kΩ and have corresponding averages and shaded standard uncertainties in green and pink, respectively. In general, the error from the TΩ meter drops as measured resistance decreases (the as-stated measurement reading errors are 0.1 % at 10 GΩ and 0.5 % beyond TΩ levels).

With the TΩ meter confirmed to operate with EG-based QHARS devices, one may now attempt to measure each of the three cross-square configurations (distinct by the recursion factor $M$). Starting with the first recursion ($M = 1$), one may use the lock-in amplifier since the predicted value is about 55.81 MΩ. Figure 3 (a) shows the resulting two-terminal magnetoresistance measurement across the main source and drain of the QHARS device while its two single Hall elements in the central area are grounded. At high $B$-fields, the magnetoresistance approaches about 54.7 MΩ and accounts for the input impedance of the circuit and a current of about 100 nA. The lower right inset of Fig. 3 (a) shows the topological drawing with a color scheme matching Fig. 1 (a) and the lower left inset shows the light blue wirebonding necessary to obtain this cross-square configuration. Another observation that can be made is that there is anticipated variation in the low-field resistance given the number of elements and the reasonable expectation that not all of them would behave precisely the same.

For the second ($M = 2$) and third ($M = 3$) recursions, one must use the TΩ meter for its high voltages and ability to measure resistance at necessary levels. Three different voltages were used to measure the second recursion: 10 V, 50 V, and 100 V. Though there is an advantage in using higher current for metrology, one must also consider the heating effects on the device and ensure that the wet cryogen system remains below 2.17 K to fully utilize the cooling power of the superfluid liquid helium, an important element for metrological purposes [38]. The time-dependent sample temperature is plotted in Fig. 3 (b) during the measurement campaign with the TΩ meter to show the effects of the application of high voltage. The magenta and lavender shaded regions indicate the times when 50 V and 100 V were applied, respectively.

The results of the measurement campaign are shown in Fig. 3 (c), with TΩ meter data on the QHARS device in its $M = 2$ iteration of the cross-square configuration (with wiring shown in (d)) suggest that higher voltages remain advantageous to precision. The expected value of 27.61 GΩ is shown as a dotted gray line, and the navy line describes the average of each full measurement campaign, along with the gold shaded area being the standard uncertainty. As expected, higher voltages translate to higher precision of the overall measurement.

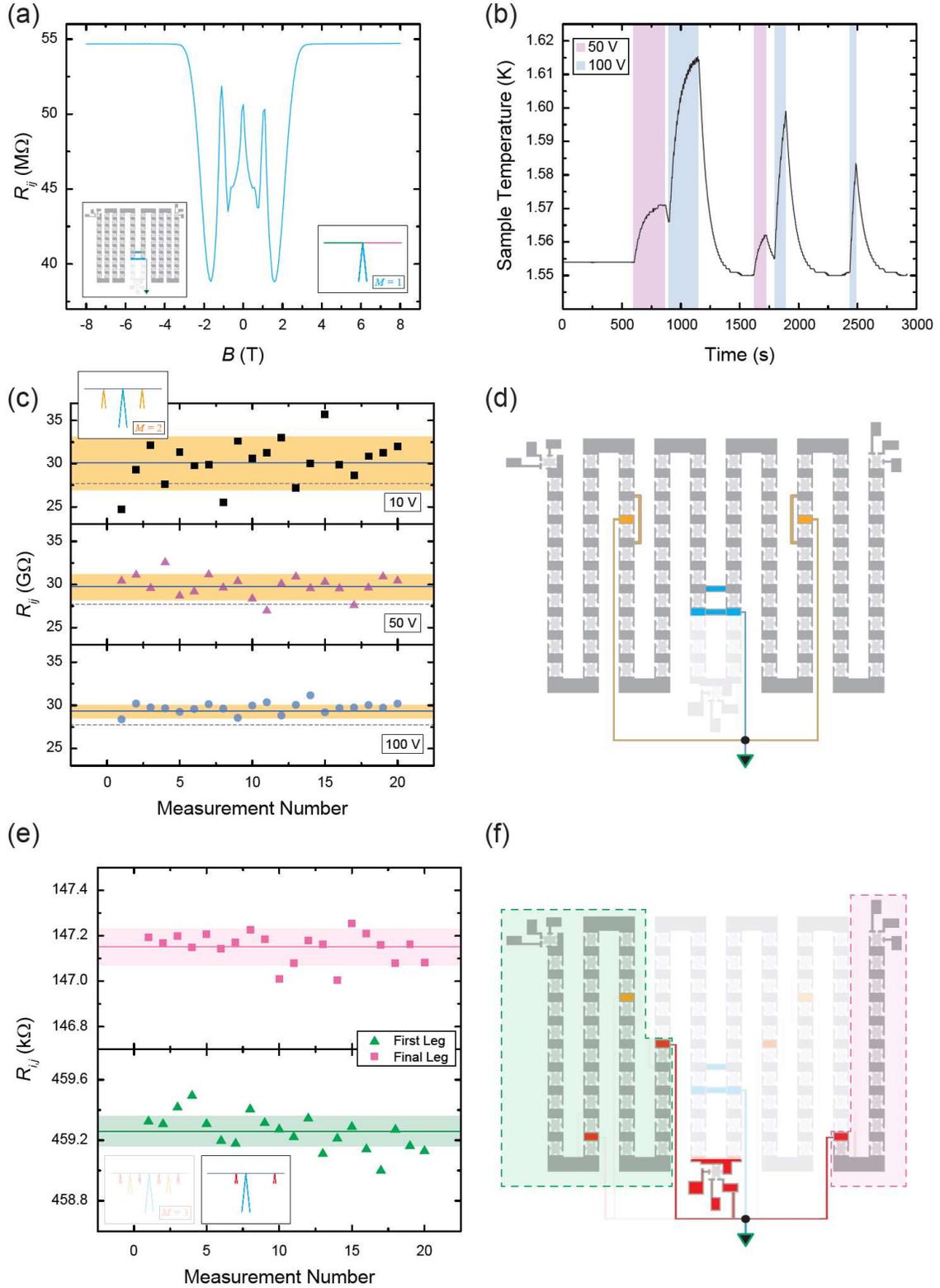

FIG. 3. (a) Two-terminal, lock-in amplifier magnetoresistance measurements were performed on the QHARS device in its $M = 1$ iteration of the cross-square configuration (where only the light blue wiring applies as shown in the lower left inset) and account for the input impedance of the circuit. The lower right inset shows the topological drawing with a color scheme matching Fig. 1 (a). (b) The time-dependent sample temperature is plotted during the course of the T$\Omega$ meter measurements to show the effects of the application of high voltage. The magenta and lavender shaded regions indicate the times when 50 V and 100 V were applied, respectively. (c) T$\Omega$ meter measurements were performed on the QHARS device in its $M = 2$ iteration of the cross-square configuration. Each graph shows a dotted

gray line indicating the calculated and simulated value of the QHARS device in the corresponding configuration, and the navy line describes the average of the full measurement campaign, along with the gold shaded area being the standard uncertainty. (d) Schematic where the light blue and orange wiring applies as shown. (e) TΩ meter measurements were performed on the QHARS device in its $M = 3$ iteration of the cross-square configuration. The solid lines and the shaded areas indicate the average and standard uncertainty of the measurement set, respectively. (f) Corresponding schematic where the light blue, orange, and red wiring applies as shown. The full *effective* high resistance was not measurable, and to verify the continued functioning of the QHARS device after application of 100 V and 500 V for prolonged periods, two distinct branches were measured per the green and pink regions at 10 V. (c) shows a dotted gray line indicating the expected value of the QHARS device subarrays.

The final configuration, with the QHARS device in its $M = 3$ iteration of the cross-square configuration (see Fig. 3 (e) and (f)), had a predicted value of 317.95 TΩ and the measurement was attempted, but the TΩ meter was unable to stabilize current flow even with the use of voltages as high as 500 V. Essentially, measuring this high a resistance would require a custom probe with exceptional electrical insulation beyond the specifications of a conventional wet cryogenic system. Despite the difficulty in obtaining this high resistance measurement, a secondary verification that was valuable to conduct was that of the continued functioning of the QHARS device after applying 100 V and 500 V for prolonged periods. To that end, in Fig. 3 (e), two distinct branches were measured with the TΩ meter and colored as green and pink regions. Each graph shows solid lines and the shaded areas indicating the average and standard uncertainty of the measurement set, respectively. The expected value of the QHARS device subarrays were about 458.2 kΩ (35.5 elements) and 148.4 kΩ (11.5 elements) for green and pink, respectively. These values being within less than 1 % of the expected is an outcome that remains promising for future QHARS devices that may need to perform at 500 V.

All in all, this work seeks to validate the mathematical framework of the cross-square, and more generally, star-mesh, transformation as a means to access very high effective quantized resistances at the GΩ level and beyond. The framework is necessary since simply building Hall elements in series ultimately constrains the upper bound of a QHARS device output. By using a maximum of 94 elements, it was shown that one can replace a QHARS device that uses hundreds of Hall elements in series in the case of accessing a quantized resistance on the order of 10 MΩ. By using pseudofractal-like recursions of the cross-square configuration, one can measure high quantized resistances until the measurement ceiling caused by resistance leakage of the conventional cryogenic system being used. Additionally, it was shown that the EG-based QHARS device design can withstand 500 V, as shown by post-exposure quantization measured at 10 V. For uses in metrology at the TΩ level and above, custom probes with exceptional electrical insulation are required, as would be more specialized bridges, like dual source bridges, to precisely determine the quality of device quantization, but based on the data presented herein, reaching such resistances via QHARS devices is likely to be reached.

## ACKNOWLEDGMENTS

NTMT and AFR designed the experiment and, with MM, WCL, collected transport data. VOJ and NTMT fabricated devices. DSS and AFR provided theoretical and mathematical support. DBN, CAR, MO, CTL, and DGJ provided general project oversight and guidance. The manuscript was written with contributions from all authors. The authors thank Y. Yang and A. R. Panna for fruitful discussions and thank L. Chao, F. Fei, and E. C. Benck, for assistance with the internal NIST review process. The authors declare no competing interest. All data are available upon reasonable request to the authors.

Commercial equipment, instruments, and materials are identified in this paper in order to specify the experimental procedure adequately. Such identification is not intended to imply recommendation or endorsement by the National Institute of Standards and Technology or the United States government, nor is it intended to imply that the materials or equipment identified are necessarily the best available for the purpose. Work presented herein was performed, for a subset of the authors, as part of their official duties for the United States Government. Funding is hence appropriated by the United States Congress directly.